\documentclass{PoS}
\usepackage{amssymb,amsmath}

\newcommand{\td}[3]{\frac{d^{#3} #1}{d {#2}^{#3}}} 

\renewcommand{\bar}[1]{\ensuremath{\overline{#1}}}

\title{Probing dark matter via neutrino-gamma-ray correlations}

\ShortTitle{Probing dark matter via neutrino-gamma-ray correlations}

\author{\speaker{Geoff Beck}\\
	School of Physics, University of the Witwatersrand, Private Bag 3, WITS-2050, Johannesburg, South Africa\\
	E-mail: \email{geoffrey.beck@wits.ac.za}}

\abstract{The nature of dark matter is one of the most pressing questions in modern cosmology. Much work has been focussed in the past upon probing potential particle dark matter via gamma-rays resulting from its annihilation or decay. These processs are dominated by the decay of pions and thus have associated neutrino fluxes. Despite this, neutrino observations have been poor in thir ability to constrain the properties of hypothetical dark matter particles due to a lack of sensitivity. Since the gamma-ray and neutrino emissions from WIMP dark matter are expected to be correlated it becomes possible to infer an associated neutrino flux to accompany any gamma-ray flux that might be attributed to dark matter. In this work we will show that it is possible to derive superior and novel constraints, particularly on leptophilic and high mass WIMP models, with this approach. This is particularly relevant in the face of leptonic-related excesses observed in both the worlds of particle and astrophysics.}

\FullConference{7th Annual Conference on High Energy Astrophysics in Southern Africa - HEASA2019\\
		28 - 30 August 2019\\
		Swakopmund, Namibia}

\begin{document}
	
\section{Introduction}
The nature and properties of Dark Matter (DM), beyond simple gravitational ones, remains a persistent anomaly in the current pictures of cosmology and particle physics. Indirect searches have made progress on a number of fronts in the past decades, namely gamma-rays~\cite{Fermidwarves2015,Fermidwarves2016,HESS:2016ygi,FermiSMC2016} and the emerging fields of neutrino astronomy~\cite{Aartsen:2018mxl,Albert:2016emp} and the use of radio astronomy in indirect DM detection~\cite{Colafrancesco2006,Colafrancesco2007,gsp2015,gs2016,chan2017,regis2017,Beck2019dsph,beckm312019}. At the same time multiple controversial excesses have been observed by various cosmic-ray experiments~\cite{ams2-antiprotons-aguilar2016,pamela-cr-spectrum2011,dampe-ambrosi2017}. These leptonic astrophysical excesses have now begun to be joined by leptonic anomalies in Large Hadron Collider (LHC)~\cite{madala-v2-1,madala-v2-2,muong-2} particularly associated the Madala hypothesis which adds several scalar bosons to account for LHC anomalies~\cite{madala1,madala2,madala-v2-1,madala-v2-2}. These persistent, often potentially DM associated, leptonic excesses point towards a need to better probe leptophilic dark matter models rather than the traditionally considered supersymmetric WIMP annihilation channels. Notably, the leptonic channel constraints from gamma-ray experiments~\cite{Fermidwarves2016,HESS:2016ygi} are considerably weaker than $b$ quarks, especially in the limit of large masses which are especially relevant to various electron-positron excesses~\cite{dampe-ambrosi2017,dampedm1,dampedm2,pamela-cr-spectrum2011}.

This work presents a novel method of obtaining stronger limits on leptophilic annihilation using gamma-ray data but translating this into a neutrino flux, as the leptophilic DM channels have more strongly peaked neutrino distributions than their gamma-ray counterparts. This translation makes use of work done in \cite{Celli:2016uon}, where the authors derive a means of finding a neutrino flux from a hadronic gamma-ray flux. This will be applicable to any higher-energy gamma-ray emissions that could be attributed to DM, as DM-associated gamma-ray emissions are dominated by hadronic processes like pion decay. The limits are then derived by comparing an inferred neutrino flux to that predicted from a target DM halo. In this work we demonstrate that the resulting limits from HESS data on the galactic centre~\cite{H.E.S.S.:2018zkf,Aharonian_2009} produce results that can be as much as an order of magnitude stronger than HESS galactic centre dark matter limits~\cite{HESS:2016ygi} on leptophilic channels for WIMP masses above 200 GeV. In particular, the $\mu$, $\tau$, $W$ channels produce results that are either close to or below the thermal relic level over the whole mass range from 200 GeV to 1 TeV when using data from the HESS galactic plane survey~\cite{H.E.S.S.:2018zkf}.

This work is structured as follows: the dark matter neutrino emission formalism is detailed in section~\ref{sec:dm}, the inference from gamma-ray to neutrino fluxes is explained in section~\ref{sec:corr}. The gamma-ray fluxes being used are elaborated on in section~\ref{sec:gamma}, results are presented in section~\ref{sec:results} and discussed in \ref{sec:conc}.

\section{Dark matter annihilation and neutrinos}
\label{sec:dm}
The source function for neutrinos from WIMP annihilations is defined as
\begin{equation}
Q_\nu (r,E) = \frac{1}{2}\langle \sigma V\rangle \sum\limits_{f}^{} \td{N^f_\nu}{E}{} B_f \left(\frac{\rho_{\chi}(r)}{m_{\chi}}\right)^2 \; ,
\end{equation}
where $\langle \sigma V\rangle$ is the velocity averaged annihilation cross-section, $f$ represents a given standard model state produced directly from annihilation (annihilation channel), $\td{N^f_\nu}{E}{}$ is the number of neutrinos per unit energy per annihilation which are found following \cite{ppdmcb1,ppdmcb2}, $B_f$ is the $f$ branching ratio, and finally $\rho_{\chi}$ and $m_{\chi}$ are the WIMP density and mass respectively. Here $E$ is the neutrino energy and $r$ is the distance from the centre of the host DM halo.
The resulting received flux on Earth will be taken to be
\begin{equation}
S_{\nu} (E) = \int_0^r d^3r^{\prime} \, \frac{Q_{\nu}(E,r)}{4\pi \left(D_L + r^{\prime}\right)^2} \; ,
\end{equation}
where $D_L$ is the luminosity distance of the halo centre.
In practice the the only $r$-dependent part of the flux is $\int_0^r d^3r^{\prime} \, \left(\frac{\rho_{\chi}(r)}{m_{\chi}}\right)^2\frac{1}{4\pi \left(D_L + r^{\prime}\right)^2}$ We will represent this with a $J$-factor defined for a given radius/angular-radius. In our case we make use to two $J$-factors, one within $0.1^\circ$ of the galactic centre and the other from $0.1^\circ$ to $1^\circ$. These were calculated assuming a Navarro-Frenk-White (NFW) halo density profile~\cite{nfw1996} with a characteristic scale of $20$ kpc and normalised to $0.3$ GeV cm$^{-3}$ in the solar neighbourhood.

\section{Neutrino-gamma correlations}
\label{sec:corr}
We use the formalism presented in \cite{Celli:2016uon} to compute a muon neutrino flux given a gamma-ray flux. This functions under the assumption of hadronic gamma-ray emissions and includes kaon, pion, and muon decay contributions. The required assumptions will not be problematic as we will only compare the inferred neutrino fluxes to those calculated from hadronicly dominated processes associated with WIMP annihilation. The calculation is performed as follows~\cite{Celli:2016uon}
\begin{equation}
\phi_{\nu\bar{\nu}} (E) = \alpha_\pi \phi_\gamma\left(\frac{E}{1-r_\pi}\right) + \alpha_k \phi_\gamma\left(\frac{E}{1-r_k}\right) + \int^{1}_{0} \left(K_{\nu} (x) + K_{\bar{\nu}}(x)\right) \phi_\gamma\left(\frac{E}{x}\right)  \; , \label{eq:corr}
\end{equation}
where $\phi_\gamma(E)$ is the gamma-ray flux at energy $E$, $\alpha_\pi = 0.658$, $\alpha_k = 0.022$, $r_k = \frac{m_\mu}{m_k}$, and $r_\pi = \frac{m_\mu}{m_\pi}$. The kernel functions $K$ are defined via
\[ K_\nu(x) = \begin{cases}
	x^2 \left(15.34 - 28.93 x\right) & 0 < x \leq r_k \\ 
	0.0165 + 0.1193 x + 3.747 x^2 - 3.981 x^3 & r_k < x < r_\pi \\ 
	(1 - x)^2 (-0.6698 + 6.588 x) & r_\pi \leq x < 1
\end{cases}\]
and
\[ K_{\bar{\nu}}(x) = \begin{cases}
x^2 \left(18.48 - 25.33 x\right) & 0 < x \leq r_k \\ 
0.0251 + 0.0826 x + 3.697 x^2 - 3.548 x^3 & r_k < x < r_\pi \\ 
(1 - x)^2 (0.0351 + 5.864 x) & r_\pi \leq x < 1
\end{cases}\]

\section{Gamma-ray fluxes}
\label{sec:gamma}
We make use of the power-law, with cut-off, fitted flux found for the central galactic source in \cite{Aharonian_2009} within $0.11^\circ$ where we note that $13\%$ of the flux in this region was found to be diffuse in origin. We model the power-law over the full range of the observations in \cite{Aharonian_2009} from 160 GeV to 70 TeV.

We also make use of an integrated flux found in annulus from $0.1^\circ$ to $1^\circ$ around the galactic centre using the HESS galactic plane survey~\cite{H.E.S.S.:2018zkf} (in particular the flux maps from the online material). This region was chosen to exclude the galactic centre source but contain some of densest regions of the DM halo. We further inferred a differential spectrum by normalising a power-law with slope $2.3$ (following the flux map modelling used in \cite{H.E.S.S.:2018zkf}) to the integrated flux over the energy range of $1$ to $100$ TeV reported in \cite{H.E.S.S.:2018zkf}. Note that in the power-law case we also extrapolate down to $160$ GeV, remaining well within the operating energy band of HESS~\cite{AHARONIAN1997369,Aharonian_2009}, in order to match the energy range of \cite{Aharonian_2009}.

\section{Results}
\label{sec:results}
Here we present $95$\% confidence interval upper limits on $\langle \sigma V\rangle$ derived by comparing the inferred neutrino flux to one predicted for the galactic DM halo within the observed regions. 

\begin{figure}[ht!]
	\centering
	\resizebox{0.49\textwidth}{!}{\includegraphics{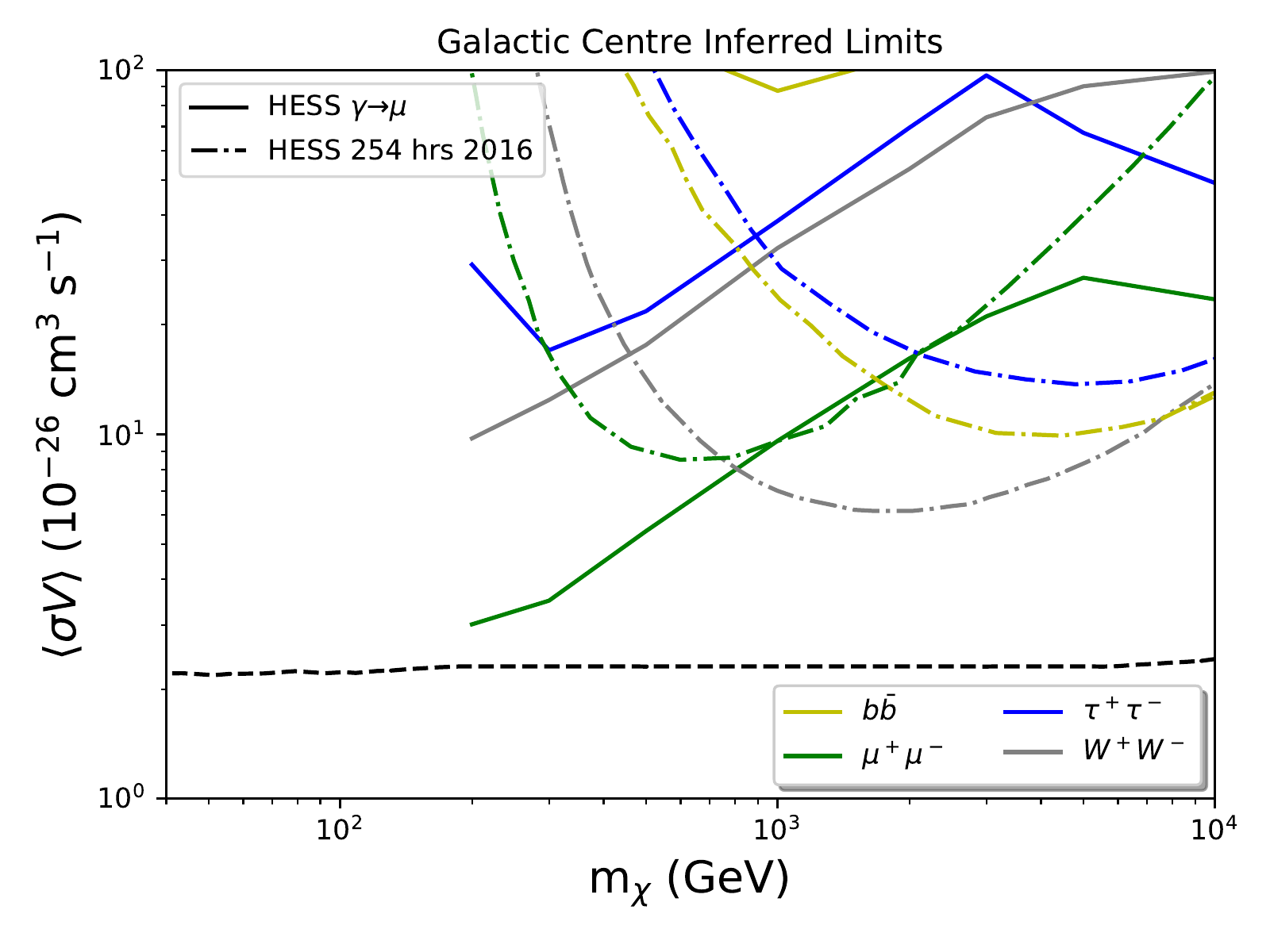}}
	\resizebox{0.49\textwidth}{!}{\includegraphics{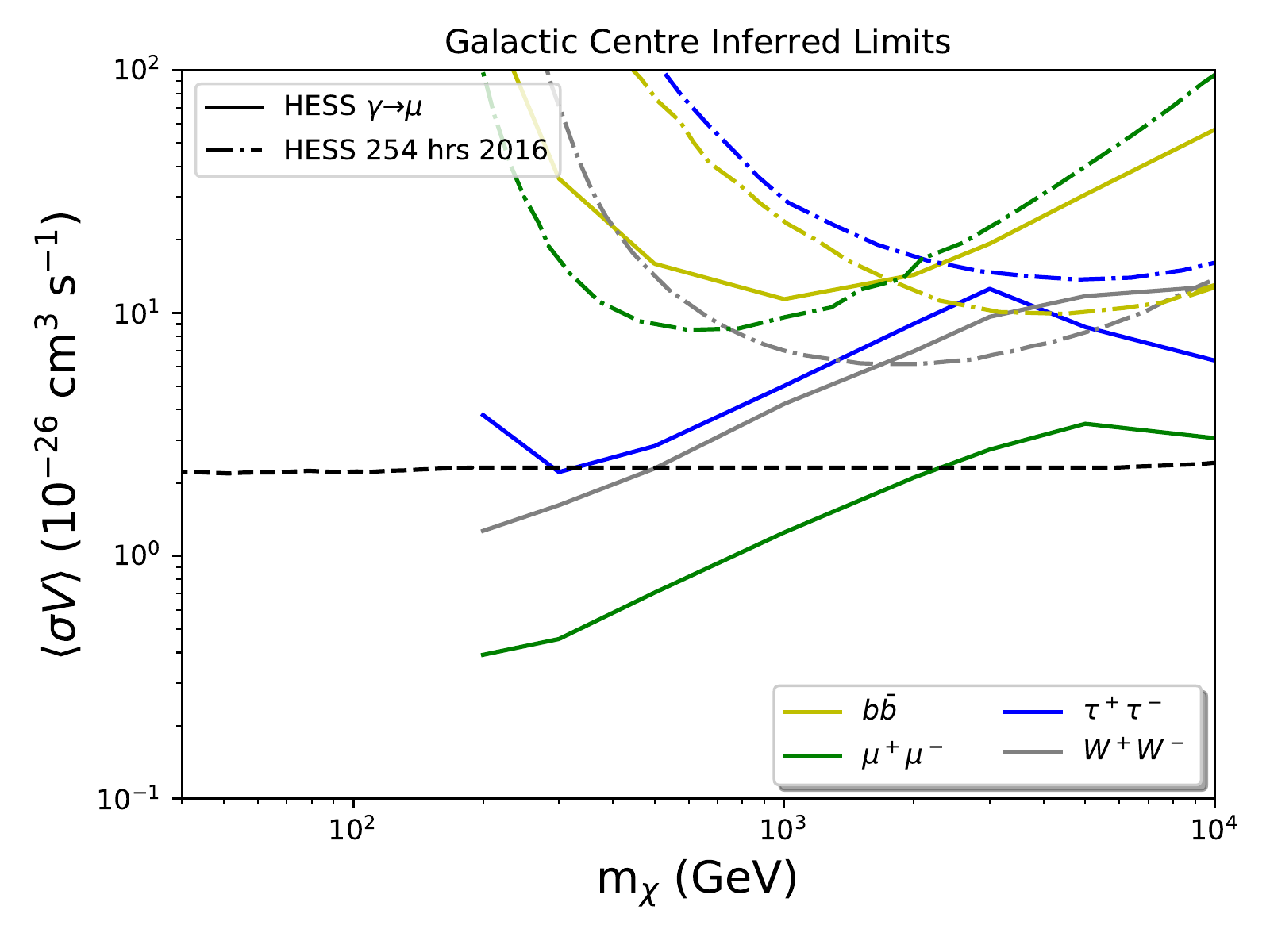}}
	\caption{Limits derived by inferring neutrino flux from GC central (within $0.11^\circ$) source gamma-ray flux taken from \cite{Aharonian_2009}. The black dashed line shows the thermal relic cross-section~\cite{steigman2012}. Left: total flux used. Right: 13\% diffuse flux used.} 
	\label{fig:hess-sgrA}
\end{figure}

In Figure~\ref{fig:hess-sgrA} we display results that make use of the gamma-ray flux found by HESS within $0.11^{\circ}$ of the galactic centre, the right panel in particular displays the results when considering the diffuse flux only. This is converted into a neutrino flux following Eq.~(\ref{eq:corr}) and compared to the predicted muon neutrino flux using a $J$-factor within $0.1^\circ$ for a variety of annihilation channels. Due to the range of the HESS data this method cannot probe below WIMP masses of 200 GeV, however, in lepton-related channels ($W^+W^-$, $\mu^+\mu^-$,$\tau^+\tau^-$) the results exceed those from the HESS study of the galactic centre~\cite{HESS:2016ygi} using 10 years of data. In particular, when the total flux is used, the limits are better than or competitive with \cite{HESS:2016ygi} when $200 < m_\chi < 500$ GeV for $W^+W^-$, when $200 < m_\chi < 900$ GeV for $\tau^+\tau^-$, and from $200$ GeV to $10$ TeV in the case of $\mu^+\mu^-$. When only the diffuse flux is considered the limits improve greatly. The muon channel is at least an order of magnitude better than \cite{HESS:2016ygi} for all studied masses (and reach below the thermal relic level up to $3$ TeV masses), for $\tau^+\tau^-$ this is true below a few TeV, and $W^+W^-$ limits are superior in the range $200 < m_\chi 1000$ GeV. The reason for this is that these particular annihilation channels produce flatter gamma-ray distributions and more pronouncedly peaked neutrino spectra. The muon channel benefits most as the neutrinos studied are of the muon flavour. The change in magnitude between the left and right panels of Fig.~\ref{fig:hess-sgrA} are expected as the diffuse flux is only $13$\% of the total~\cite{Aharonian_2009}.

\begin{figure}[ht!]
	\centering
	\resizebox{0.49\textwidth}{!}{\includegraphics{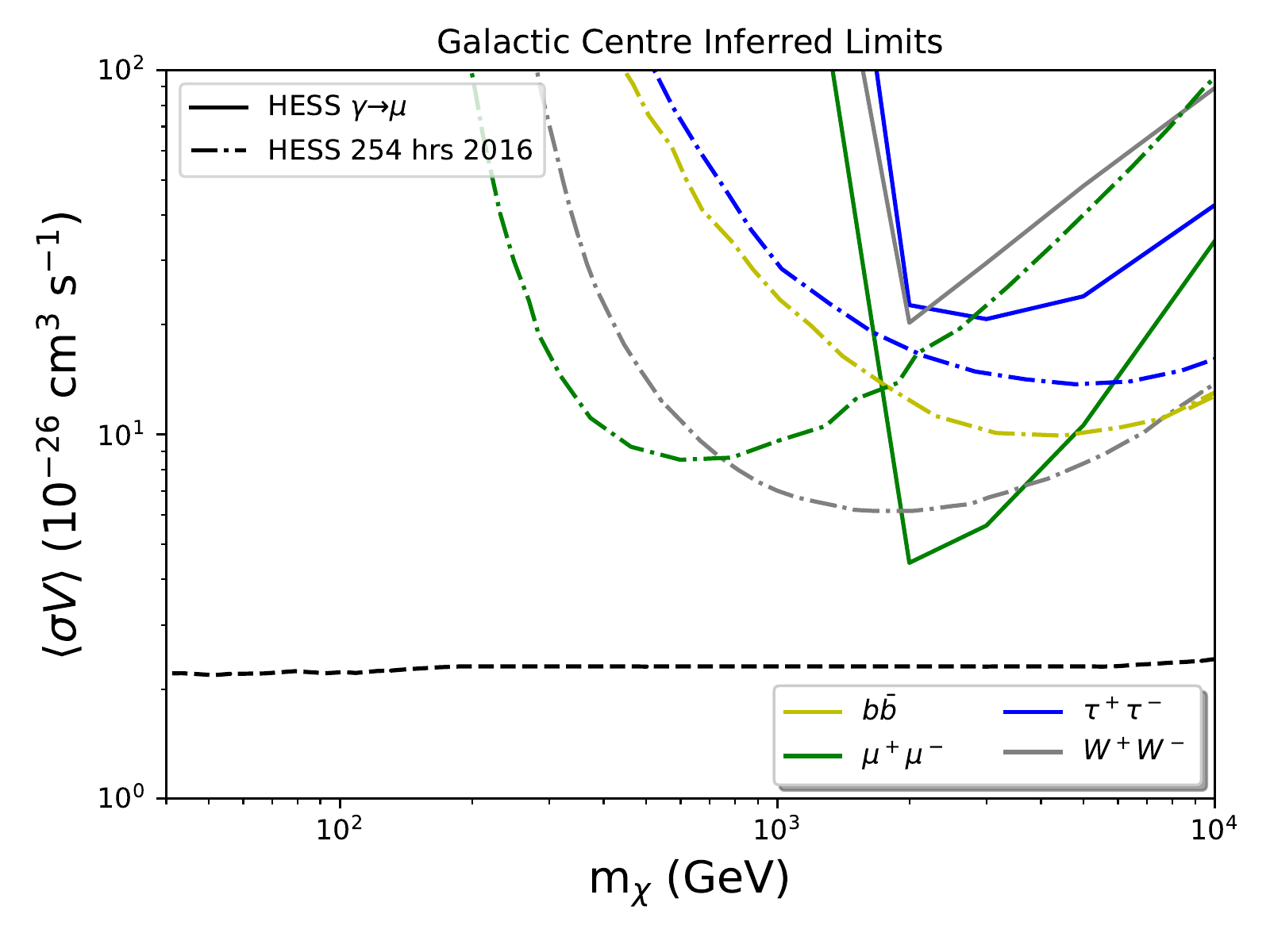}}
	\resizebox{0.49\textwidth}{!}{\includegraphics{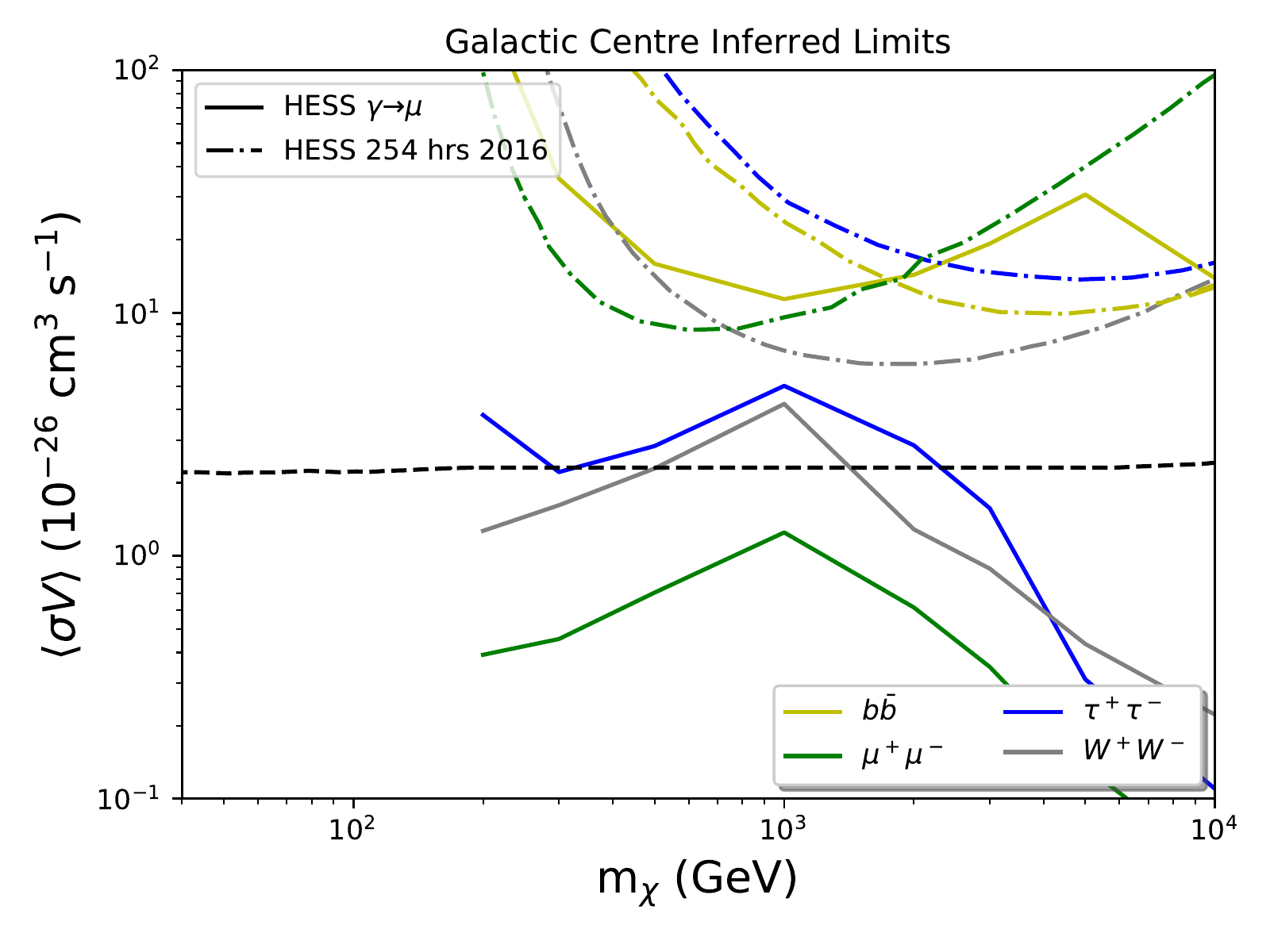}}
	\caption{Limits derived by inferring neutrino flux from HESS galactic plane survey gamma-ray flux within region from $0.1^\circ$  to $1^\circ$ taken from the flux maps described by \cite{H.E.S.S.:2018zkf}. The black dashed line shows the thermal relic cross-section~\cite{steigman2012}. Left: integrated flux from map used. Right: differential flux calculated for power-law slope $2.3$ used. }
	\label{fig:hgpc}
\end{figure}

In Fig.~\ref{fig:hgpc} the results derived from the HESS galactic plane survey data are displayed~\cite{H.E.S.S.:2018zkf}. Here we have selected an observations region in the annulus $0.1^\circ$ to $1^\circ$ around the galactic centre. We compare to two cases, one that uses the integrated flux taken directly from the HESS maps, the other where we normalise a power-law with slope $2.3$ to match this integrated flux within an energy range of $1$ to $100$ TeV (these parameters reflect the source modelling used in \cite{H.E.S.S.:2018zkf} for the flux map generation). The non-integrated case is of importance as the shape of DM-produced spectra is one of the most useful properties in their comparison to more mundane astrophysical processes. In the integrated flux case (left hand panel of \ref{fig:hgpc}) we see that only very limited gains can be made over the existing HESS galactic centre results. This is mostly confined to large mass WIMPs as the integrated flux is only taken above $1$ TeV. Despite this limitation we find that the muon channel shows substantial gains for masses above $1$ TeV. When a differential flux is used instead (right panel of Fig.~\ref{fig:hgpc}) we see very substantial gains across all the studied channels. What is most notable is the shape of the constraint curves, which are similar to those from Fig.~\ref{fig:hess-sgrA} at low masses but above $1$ TeV have a negative slope making them unusually powerful for the study of high-mass WIMPs which are usually difficult to probe. Importantly, even the $b$ quark channel has superior limits from this method for masses below $1$ TeV. In the leptonic $\mu$, $\tau$, and associated $W$ channels the results are superior at all masses above $200$ GeV and either close to or below the thermal relic level over the whole mass range.

\section{Discussion and conclusions}
\label{sec:conc}
The existence of multiple, albeit controversial, cosmic-ray excesses associated with leptons as well as the emerging LHC excesses associated with the Madala hypothesis make it plain that the leptonic sector is becoming a rich hunting ground for exotic physics. In this regard it is of special interest in the hunt for a DM candidate. However, gamma-ray indirect probes have historically been at their weakest when studying leptophilic annihilation/decay channels. This work has presented a method where gamma-ray data can be used to produce far more stringent limits on leptophilic WIMP models by inferring a neutrino flux and comparing this to the DM predictions (as these annihilation channels produce more peaked neutrino than gamma-ray spectra). 

We have demonstrated that superior limits to the HESS galactic centre results with 10 years of data can be obtained when making use of both the diffuse flux within $0.11^\circ$ of the galactic centre from~\cite{Aharonian_2009} as well as a power-law gamma-ray spectrum fitted to the annulus between $0.1^\circ$ and $1^\circ$ drawn from the HESS galactic plane survey~\cite{H.E.S.S.:2018zkf}. Importantly, the obtained results were up to an order of magnitude better for the heavy lepton channels as well as the $W$ boson case from 200 GeV to 10 TeV. Further investigation will go into supplementing these reslts with gamma-ray fluxes from lower-energy instruments like Fermi-LAT as well as ensuring the robustness of the neutrino flux inference. It may also be of importance to determine whether a similar inference can be made for electron neutrinos, as the electron annihilation channel has historically weak gamma-ray limits, similar or weaker than the muon case.

\bibliographystyle{JHEP}
\bibliography{heasa2019,dm_indirect,dm_indirect_general,gbeck,madala_all,instruments_all,dsph_all}

\end{document}